\pdfoutput=1
\documentclass[aps,pre,preprint,a4paper,superscriptaddress]{revtex4-2}
\usepackage{amsmath,amssymb,amsfonts}
\usepackage{graphicx}
\usepackage{bm}
\usepackage{color}
\usepackage{hyperref}
\usepackage{xcolor}
\usepackage{enumitem}
\usepackage{multirow}
\usepackage{booktabs}
\usepackage{indentfirst}

\hypersetup{colorlinks=true,linkcolor=blue,urlcolor=blue,citecolor=blue}

\usepackage{geometry}
\begin{document}
	
	\title{Global Minima of the Thomson Problem in a Disk: A Molecular Dynamics Approach with Fixed Border Charges}
	
	\affiliation{Dzhelepov Laboratory of Nuclear Problems, JINR, Dubna, Russian Federation}
	\affiliation{Meshcheryakov Laboratory of Information Technologies, JINR, Dubna, Russian Federation}
	\affiliation{Dubna State University, Dubna, Russian Federation}
	
	\author{G.\,K.~Lavrov}
	\email{lavrov@jinr.ru}
	\affiliation{Dzhelepov Laboratory of Nuclear Problems, JINR, Dubna, Russian Federation}
	\affiliation{Dubna State University, Dubna, Russian Federation}
	
	\author{E.\,G.~Nikonov}
	\email{e.nikonov@jinr.ru}
	\affiliation{Meshcheryakov Laboratory of Information Technologies, JINR, Dubna, Russian Federation}
	\affiliation{Dubna State University, Dubna, Russian Federation}
	\affiliation{HSE University, Moscow, Russian Federation}
	
	\date{\today}
	
	\begin{abstract}
		We report improved global--minimum configurations for the classical Thomson problem of $N=60$,\,$61$,\,$92$, and $99$ repulsive Coulomb charges confined to a disk. By combining the quenched molecular dynamics (QMD) method with the fixed--border heuristic introduced by Amore and Zarate, we systematically obtain configurations with energies $E_{\mathrm{QMD}}(60)=2159.3584240930$, $E_{\mathrm{QMD}}(61)=2237.19264190$, $E_{\mathrm{QMD}}(92)=5358.35353314$, and $E_{\mathrm{QMD}}(99)=6254.83029083$, which improve upon the previously best--known values. For $N=60$, the Voronoi diagram of our configuration differs from the one reported earlier. The symmetries for $N=61$ ($C_{2}$) and $N=99$ ($D_{1}$) are confirmed and are consistent with the known symmetry patterns for these configurations. For $N=92$, whereas the previously reported Voronoi diagram has lower symmetry, our solution exhibits a clear $D_{1}$ (axial) symmetry of defects. The results are highly reproducible across multiple independent runs, providing strong evidence that these configurations are robust global--minimum candidates.
	\end{abstract}
	

	\maketitle
	\newpage
	
	\section{Introduction}
	The distribution of repulsive particles in confined geometries is a fundamental problem in classical physics. Closely related ordering phenomena include magnetic--field--induced Wigner solids in two--dimensional electron systems~\cite{Andrei1988}, Abrikosov vortex lattices in superconductors~\cite{Gammel1987}, charged colloidal and dusty-plasma crystals~\cite{Yethiraj2003,Thomas1994}, and correlated electronic states in moir\'e heterostructures~\cite{Andrei2020, Zhou2021}. The Thomson problem in a disk, which consists of finding the ground state of $N$ mutually repelling Coulomb charges confined to a unit disk, has attracted considerable attention due to its rich topological structure and computational complexity~\cite{Bedanov1994,Mughal2007,Cerkaski2015}. Since the seminal work of J.\,J.~Thomson~\cite{Thompson1904}, it has continued to attract considerable attention in chemistry and physics.
	
	The ground-state configurations discovered for small systems (\(N \le10^{2}\)) are in most cases characterized by discrete rotational and axial symmetries of defects, including the well-known ``magic-number'' states~\cite{Nazmitdinov2017}. Increasing the system size to \(N \sim10^{3}\) leads to the formation of dislocation scars composed of alternating five- and seven-fold disclination chains~\cite{Irvine2010,Bausch2003}. For macroscopic systems (very large \(N\)), the structure transforms into a bulk polycrystal, which is screened from the outer edge by a disordered boundary layer~\cite{Bedanov1994,Koulakov1998,Worley2006}. A key challenge in finding the global minimum is correctly identifying the number of charges that reside on the boundary, $N_b$, which scales as $N^{2/3}$~\cite{Worley2006,Amore2023}. The fixed--$N_b$ heuristic introduced by Amore and Zarate~\cite{Amore2023} is particularly useful because it significantly reduces the effective dimensionality of the search and has made it possible to obtain high--quality candidate minima for many values of $N$. Using this heuristic combined with ``Divide \& Conquer'' and ``Basin--Hopping'' methods~\cite{Wales1997}, they reported candidate global minima for various $N$, including $N=60$,\,$61$,\,$92$, and $99$.
	
	In this Letter, we demonstrate that a physically motivated approach --- quenched molecular dynamics (QMD) --- coupled with the same fixed--$N_b$ heuristic yields configurations with lower energy and, in several cases, higher symmetry. These results provide improved global--minimum candidates for $N=60,61,92,99$ and illustrate the value of combining geometric constraints with physical relaxation dynamics.
	
	\section{Problem Statement}
	We consider a system of $N$ identical classical charges confined to a disk of unit radius. The particles interact via the repulsive Coulomb potential. The total energy of the system is given by the Hamiltonian
	\begin{equation}
		H = \sum_{i=1}^{N} V(r_i) + \sum_{i<j}^{N} \frac{1}{|\mathbf{r}_i - \mathbf{r}_j|},
		\label{eq:hamiltonian}
	\end{equation}
	where $\mathbf{r}_i = (x_i, y_i)$ is the position of the $i$th charge, $r_i = |\mathbf{r}_i|$ is its distance from the center of the disk, and $V(r)$ is the confining potential. In the present work, we consider the hard--wall (disk) confinement:
	\begin{equation}
		V(r) =
		\begin{cases}
			0, & r \leq R \\
			\infty, & r > R
		\end{cases},
	\end{equation}
	with the disk radius set to $R = 1$ without loss of generality. The equilibrium configurations correspond to the global minimum of the Hamiltonian~\eqref{eq:hamiltonian} subject to the constraint $r_i \leq 1$ for all $i$.
	
	A distinctive feature of this problem is the accumulation of charges near the boundary, which leads to the formation of a distinct border layer. The number of charges residing on the boundary, $N_b$, scales as $N^{2/3}$ for large $N$~\cite{Mughal2007,Worley2006,Amore2023}. The correct identification of $N_b$ is crucial for finding the global minimum, as it significantly reduces the dimensionality of the configuration space.
	
	\section{Method}
	Our approach is based on the quenched molecular dynamics (QMD) algorithm, successfully applied in Ref.~\cite{Nazmitdinov2017}, which integrates the equations of motion for charges with a damping term. The equations of motion are:
	\begin{equation}
		m \cdot \ddot{\mathbf{r}}_i = \mathbf{F}_i - b \cdot \dot{\mathbf{r}}_i,
	\end{equation}
	where $\mathbf{F}_i = \sum_{j \neq i} (\mathbf{r}_i - \mathbf{r}_j)/|\mathbf{r}_i - \mathbf{r}_j|^3$ is the Coulomb force, and $b$ is the damping (quenching) parameter. This is analogous to a system moving in a viscous medium, and it naturally relaxes to local minima of the potential energy surface.
	
	To address the specific geometry of the disk, we fix the number of border charges $N_b$ using the heuristic from Ref.~\cite{Amore2023}:
	\begin{equation}
		N_b^{\mathrm{fit}}(N) = 2.84328 \, N^{2/3} - 0.530196 \, N^{1/3} - 2.32866,
	\end{equation}
	rounded to the nearest integer. Charges on the boundary are constrained to move tangentially, while interior charges are allowed to move freely. To avoid trapping in shallow local minima, we employ a gradual increase of the parameter $\sigma$, which defines the effective radius of the interior charges, from its initial value $\sigma=1-1/(2\sqrt{N})$ to $\sigma=1$, as suggested in Ref.~\cite{Amore2023}. This allows the system to smoothly accommodate increasing density without abrupt structural rearrangements.
	
	The algorithm is implemented in Python with Numba--accelerated force computations, allowing for efficient exploration of the energy landscape. For each value of $N$, we performed $150\,000$ time steps with a time step $\Delta t=0.001$ and a damping parameter $b=0.9$, followed by a final relaxation phase. The reproducibility of our results was verified by performing $10$ independent runs with different random initial conditions.
	
	\section{Results}
\begin{figure}[h]
		\centering
		\begin{minipage}[b]{0.22\textwidth}
			\centering
			\includegraphics[width=\textwidth]{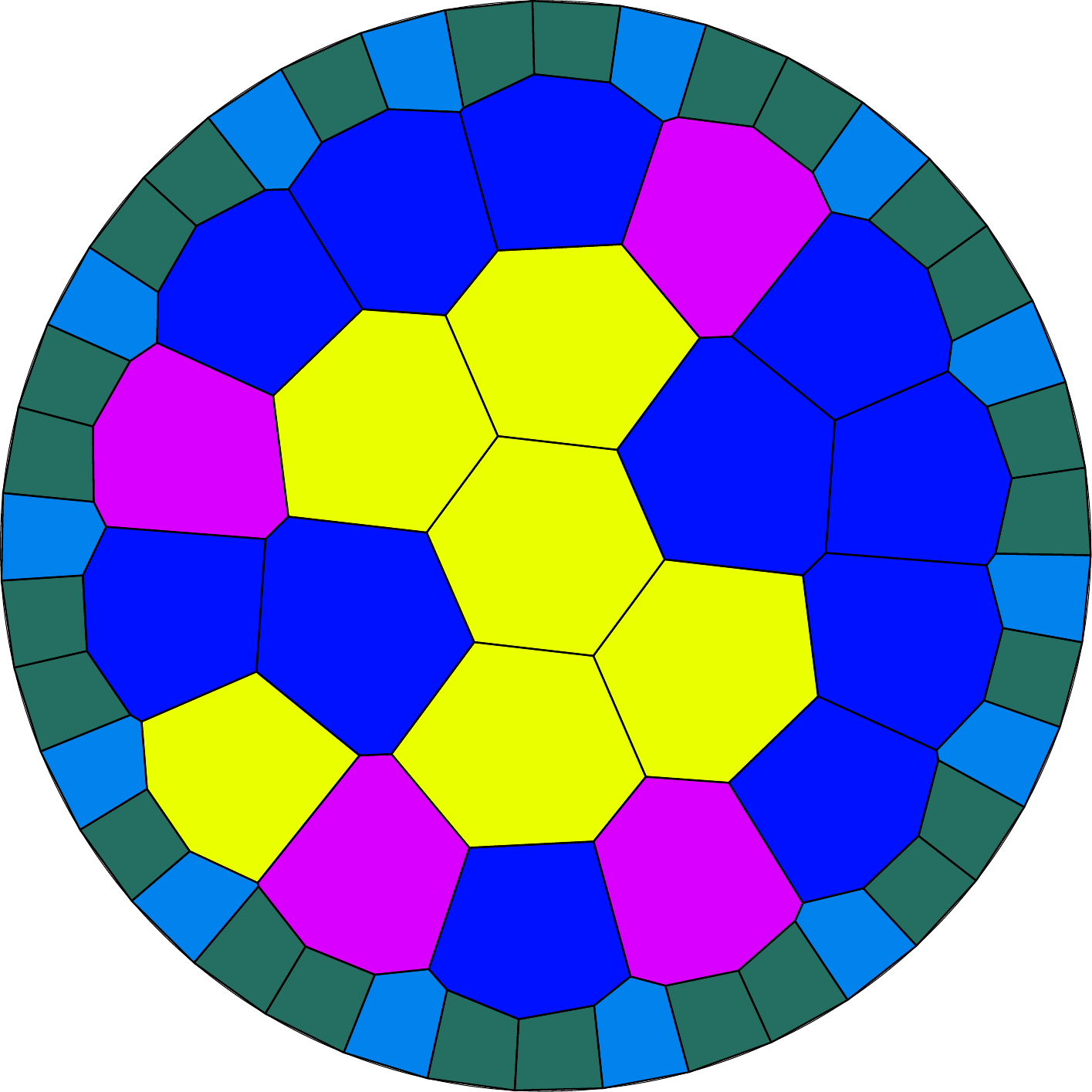}
			\footnotesize
			\textbf{(a)} $N=60$, $N_b=39$\\(---)
		\end{minipage}
		\hfill
		\begin{minipage}[b]{0.22\textwidth}
			\centering
			\includegraphics[width=\textwidth]{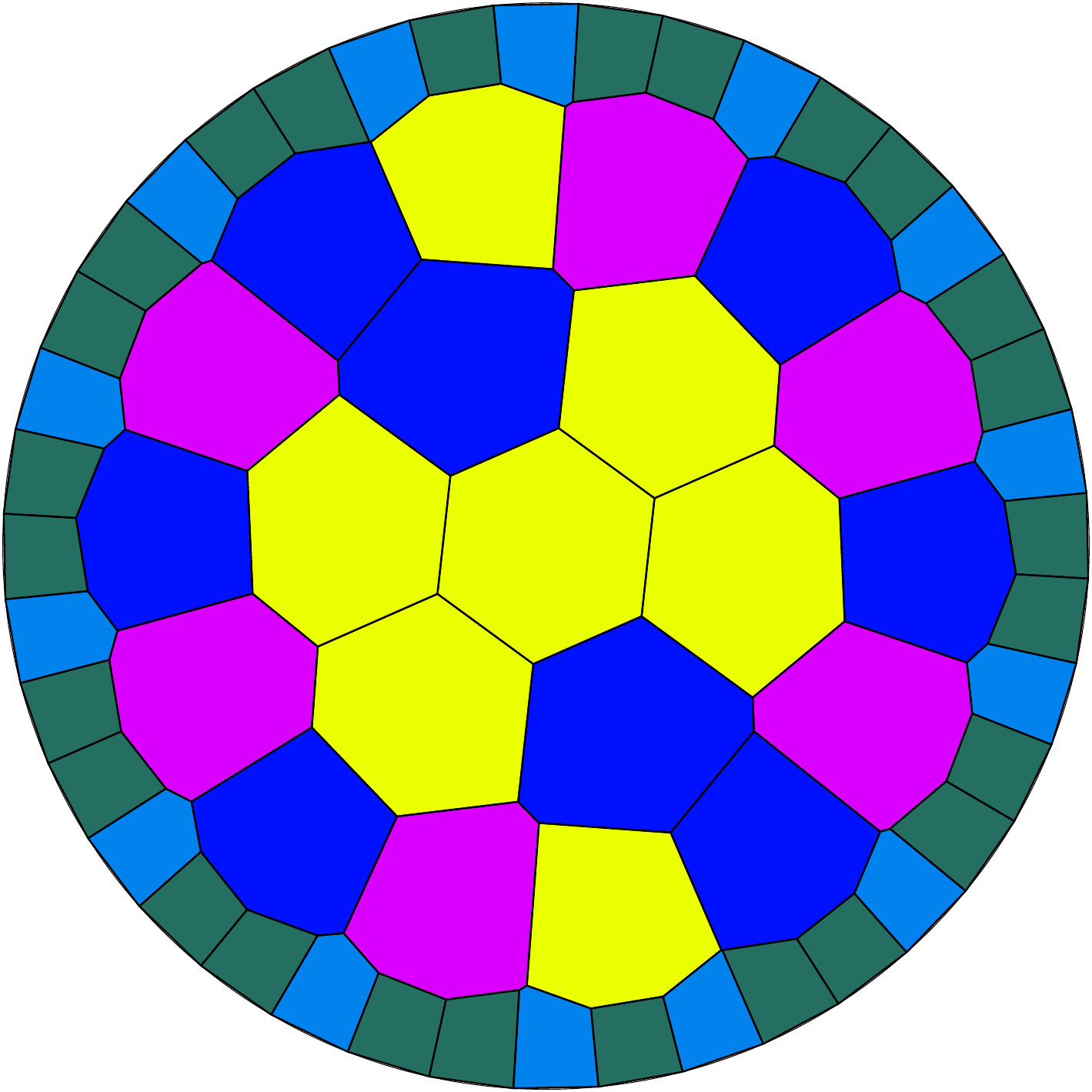}
			\footnotesize
			\textbf{(b)} $N=61$, $N_b=40$\\($C_2$)
		\end{minipage}
		\hfill
		\begin{minipage}[b]{0.22\textwidth}
			\centering
			\includegraphics[width=\textwidth]{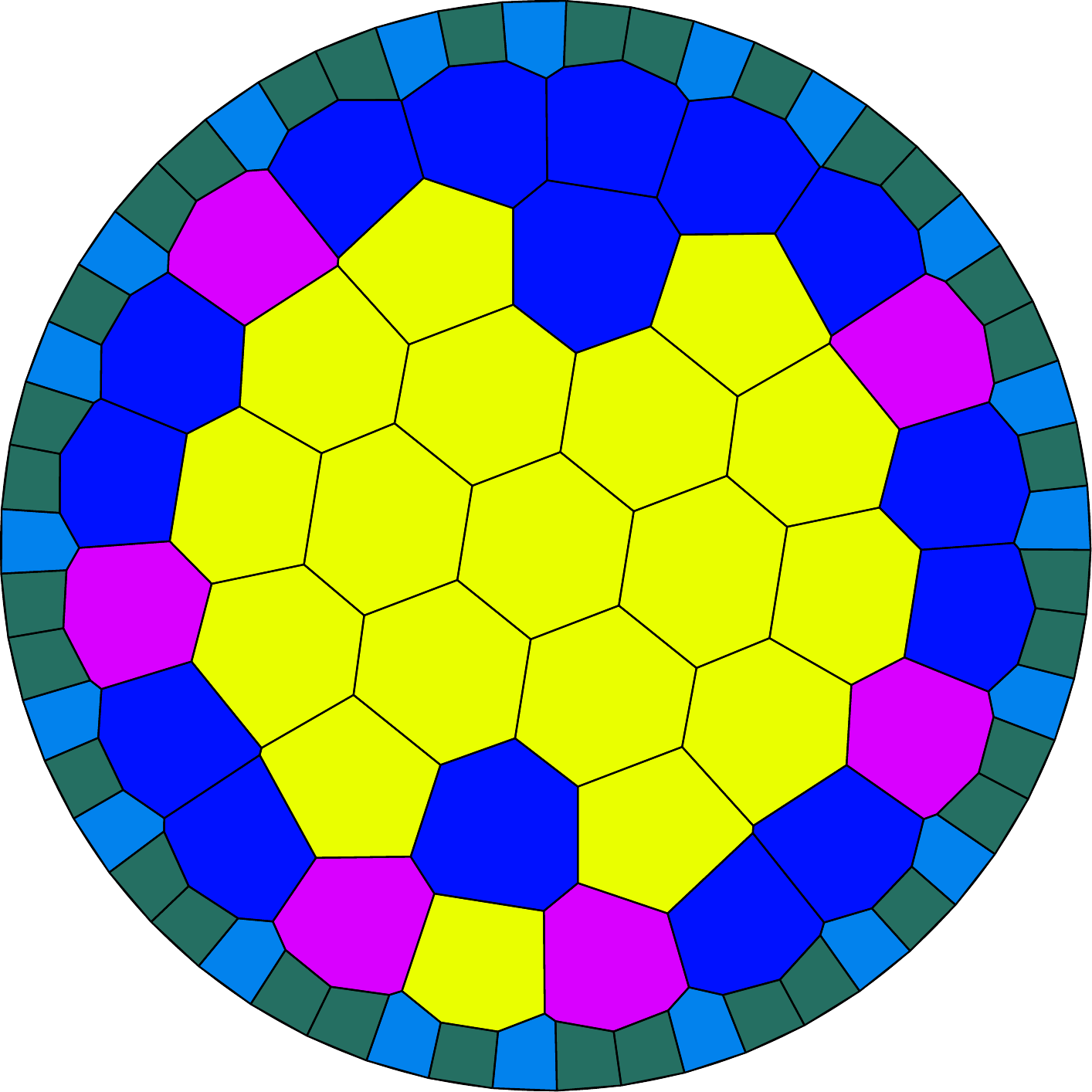}
			\footnotesize
			\textbf{(c)} $N=92$, $N_b=53$\\($D_1$)
		\end{minipage}
		\hfill
		\begin{minipage}[b]{0.22\textwidth}
			\centering
			\includegraphics[width=\textwidth]{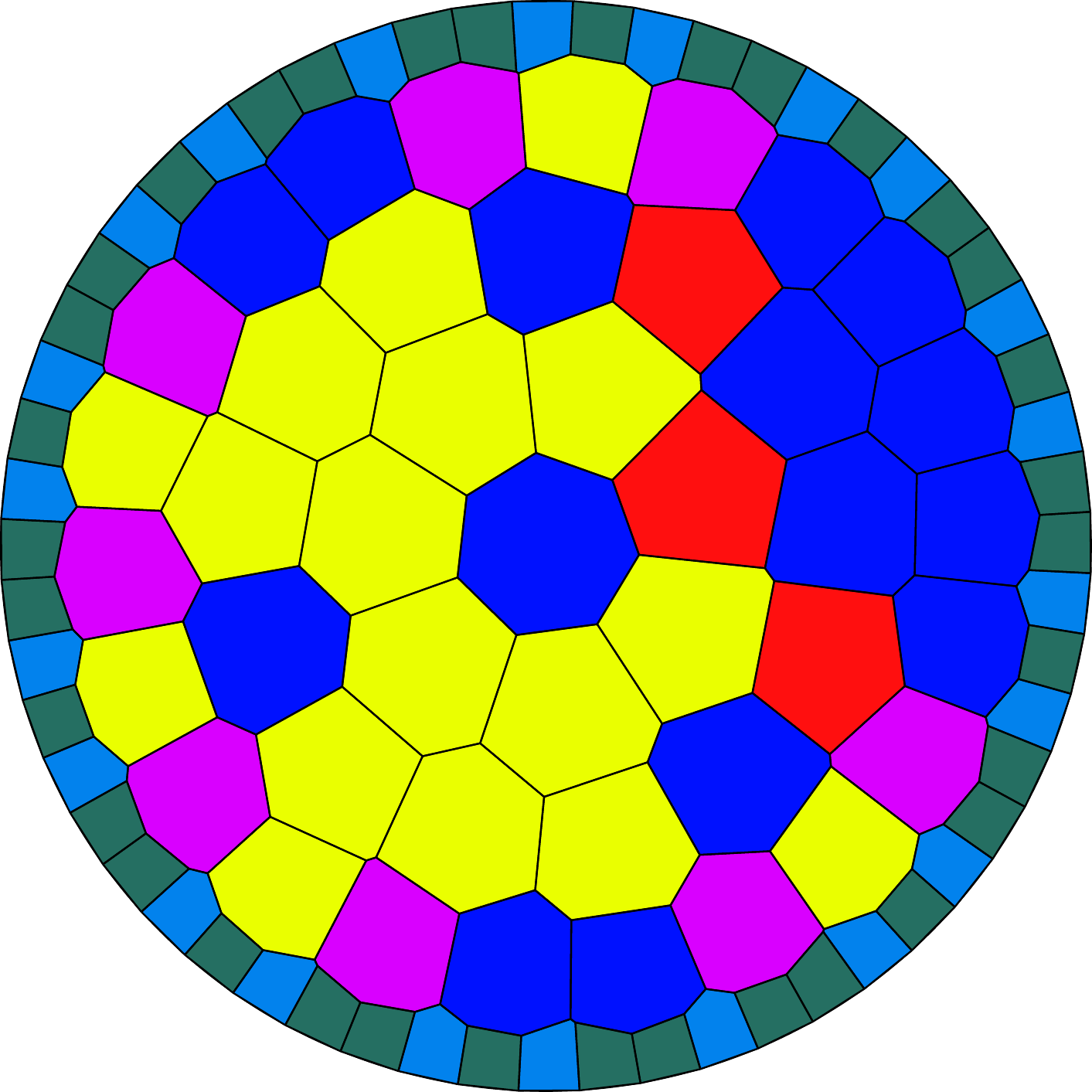}
			\footnotesize
			\textbf{(d)} $N=99$, $N_b=56$\\($D_1$)
		\end{minipage}
		\caption{Voronoi diagrams of our ground--state configurations for (a) $N=60$,\,(b) $61$,\,(c) $92$,\,and\,(d) $99$. The corresponding symmetry groups are presented in brackets. The Voronoi diagram for $N=60$ differs from the one published earlier. For $N=61$, the Voronoi diagram of our configuration is a mirror--related variant of the structure reported earlier. For $N=92$, our solution exhibits a clear $D_{1}$ symmetry, whereas the previously reported structure has lower symmetry. For $N=99$, the Voronoi diagram of our configuration coincides with the previously published one.}
		\label{fig:voronoi}
	\end{figure}

	Figure~\ref{fig:voronoi} shows the Voronoi diagrams of our ground--state configurations for $N=60$,\,$61$,\,$92$, and $99$. The configuration for $N=61$ exhibits $C_{2}$ symmetry, while the configurations for $N=92$ and $N=99$ exhibit $D_{1}$ (mirror) symmetry of defects, with the defect structures arranged symmetrically about a central axis. Interestingly, for $N=61$, the Voronoi diagram of our configuration is a mirror--related variant of the structure reported in Ref.~\cite{Amore2023SM}, while possessing a slightly lower energy; this suggests that the two structures are closely related minima and that QMD selects the more stable one. For $N=92$, our solution exhibits a clear $D_{1}$ symmetry, whereas the configuration reported in Ref.~\cite{Amore2023SM} has lower symmetry. This comparison suggests that the earlier result is likely a low--energy local minimum, while the symmetric configuration obtained here is the more stable ground--state candidate.
	
	The energies of our configurations ($E_{\mathrm{QMD}}$) are:
	\begin{align}
		E_{\mathrm{QMD}}(60) &= 2159.3584240930, \\
		E_{\mathrm{QMD}}(61) &= 2237.19264190, \\
		E_{\mathrm{QMD}}(92) &= 5358.35353314, \\
		E_{\mathrm{QMD}}(99) &= 6254.83029083.
	\end{align}
	These values are lower than those reported in Ref.~\cite{Amore2023SM}. The differences, while small, are systematic and significant: we obtained the same energies and the same symmetric structures in $10$ independent runs for each $N$, confirming the robustness of our method. Table~\ref{tab:energies} summarizes the improvements.
	
	\begin{table}[h!]
		\centering
		\caption{Comparison of energies for $N=60$,\,$61$,\,$92$, and $99$ between our configurations
			($E_{\mathrm{QMD}}$) and those reported in Ref.~\cite{Amore2023SM} ($E_{\mathrm{A-Z}}$).}
		\setlength{\tabcolsep}{5pt} 
		\begin{tabular}{ccrrr}
			\toprule[0.4pt]
			\toprule[0.4pt]
			$N$ & $N_{b}$ &
			\multicolumn{1}{c}{$E_{\mathrm{QMD}}$} &
			\multicolumn{1}{c}{$E_{\mathrm{A-Z}}$} &
			\multicolumn{1}{c}{$\Delta E$} \\
			\midrule[0.4pt]
			60 & 39 & 2159.3584240930 & 2159.3584240938 & $8.0 \times 10^{-10}$ \\
			61 & 40 & 2237.19264190\phantom{00} & 2237.19264194\phantom{00} & $4.0 \times 10^{-8\phantom{0}}$ \\
			92 & 53 & 5358.35353314\phantom{00} & 5358.35353346\phantom{00} & $3.2 \times 10^{-7\phantom{0}}$ \\
			99 & 56 & 6254.83029083\phantom{00} & 6254.83029092\phantom{00} & $9.0 \times 10^{-8\phantom{0}}$ \\
			\bottomrule[0.4pt]
		\end{tabular}
		\label{tab:energies}
	\end{table}
	
	\section{Discussion}
	The combination of QMD with fixed--$N_b$ constraints provides a powerful tool for exploring the energy landscape of confined charged systems. The damping term in QMD effectively removes kinetic energy while allowing the system to follow the steepest descent paths, avoiding many shallow local minima that can trap purely gradient--based methods. The gradual increase of the parameter $\sigma$, which defines the effective radius of the interior charges, further guides the system through the complex landscape by gradually introducing the confining potential.
	
	The $D_{1}$ symmetry of our $N=92$ configuration is a strong indicator of its global nature.	In systems with symmetric domains, the ground state often preserves the symmetry of the confining geometry, whereas lower--symmetry configurations can correspond to metastable minima. The lower--symmetry structure reported previously for $N=92$~\cite{Amore2023SM} may therefore represent such a metastable state, while the symmetric configuration found here is energetically preferred. For $N=61$ and $N=99$, the symmetries we observe ($C_{2}$ and $D_{1}$, respectively) are consistent with the expected patterns for these configurations, further supporting our approach.
	
	Our results also highlight a broader and more profound aspect of the Thomson problem: \textit{there is no single universal optimization algorithm that can reliably identify the true global minimum for arbitrary $N$}. The energy landscape of this system is extraordinarily complex, with an exponentially growing number of local minima as $N$ increases (for the Thomson problem on a surface of the sphere see Ref.~\cite{Amore2025}). Different algorithms --- whether gradient--based methods like ``truncated Newton'', stochastic methods like ``Basin--Hopping'', or physically motivated approaches like QMD --- each have their own strengths and weaknesses, and none is universally superior. The ``Basin--Hopping'' method, for instance, excels at escaping local minima through Monte Carlo moves, while QMD follows physical trajectories that naturally respect the dynamics of the system. The ``Divide \& Conquer'' strategy of Ref.~\cite{Amore2023} is an important advance that substantially reduces the effective dimensionality of the problem. Our results suggest that combining this strategy with physical relaxation dynamics can be especially productive, because the dynamics helps the system pass through shallow minima and recover highly symmetric low--energy structures.
	
	This observation suggests that the most effective strategy for finding global minima in such systems is the development of \textit{hybrid approaches} that combine the strengths of multiple algorithms. By sequentially applying different optimization techniques or by using one method to generate initial configurations for another, it may be possible to explore the energy landscape more thoroughly and consistently identify the true ground state.
	
	Furthermore, our findings motivate continued cross--checking of reported minima for the Thomson problem in a disk. The identification of lower--energy configurations for $N=61$,\,$92$, and $99$ suggests that similar improvements may exist for other values of $N$, including both smaller and larger systems. The systematic symmetry of our configurations, combined with their reproducibility across many independent runs, provides strong evidence that these are indeed robust global--minimum candidates. However, the absence of a universally applicable algorithm means that one cannot assume that any given configuration, even if obtained with a sophisticated method, is necessarily the true ground state.
	
	It is also instructive to compare our results for $N=60$, a case where the energy reported by Amore and Zarate, $E_{\mathrm{A-Z}}(60) = 2159.3584240938$, is very close to our value $E_{\mathrm{QMD}}(60) = 2159.3584240930$. Despite the seemingly negligible energy difference ($\Delta E = 8.0 \times 10^{-10}$), the corresponding Voronoi diagrams are structurally distinct. This observation further supports the conclusion that even for moderate $N$, the energy landscape contains multiple local minima with very similar energies but different topological defect structures. A physically motivated dynamics--based approach such as QMD can help discriminate between such nearly degenerate states.
	
	In this context, the QMD approach presented here offers a valuable addition to the toolkit for studying confined Coulomb systems. Its physical intuition, combined with its ability to consistently produce symmetric, low--energy configurations, makes it a promising candidate for future investigations. We suggest that a comprehensive re--evaluation of the Thomson problem in a disk, using a diverse set of complementary algorithms, is warranted and may reveal a richer structure of global minima than previously appreciated.
	
	\section{Conclusion}
	We have found improved global--minimum configurations for the Thomson problem in a disk with $N=60$,\,$61$,\,$92$, and $99$ charges, with energies $E=2159.3584240930$, $2237.19264190$, $5358.35353314$, and $6254.83029083$, respectively. The configurations exhibit the expected symmetries: $C_{2}$ for $N=61$, and $D_{1}$ for $N=92$ and $N=99$. In particular, for $N=92$, our solution exhibits a clear $D_{1}$ symmetry, whereas the previously reported structure has lower symmetry. The configurations are reproducibly obtained using a QMD algorithm with fixed border charges. Our results refine the previously reported best energies by $8.0 \times 10^{-10}$, $4.0 \times 10^{-8}$, $3.2 \times 10^{-7}$, and $9.0 \times 10^{-8}$, respectively, and demonstrate the effectiveness of QMD for studying confined Coulomb systems.
	
	More broadly, our work underscores the inherent difficulty of finding global minima in the Thomson problem and highlights the need for hybrid and multi--algorithm approaches. The current set of reported minima should be carefully cross--checked, as no single algorithm is universally capable of identifying the true ground state for arbitrary $N$. We suggest that the algorithms introduced here can be extended to larger $N$ and to more general domain geometries, opening new avenues for research in classical and soft matter physics.
	
	\section*{Data Availability}
	The data that support the findings of this article are publicly available (see the supplemental materials in Refs.~\cite{Lavrov2026SM1, Lavrov2026SM2}). The software developed and used in this study is the intellectual property of the authors and the Joint Institute for Nuclear Research and therefore cannot be distributed in the public domain or provided upon request.
	
	\section*{Acknowledgments}
	The authors are grateful to Professor R.\,G.~Nazmitdinov for valuable discussions and comments on the manuscript.
	
	\section*{FUNDING}
	This work was supported by ongoing institutional funding at the Joint Institute for Nuclear Research. No additional grants to carry out or direct this particular research were obtained.
	
	\section*{CONFLICT OF INTEREST}
	The authors declare that they have no conflicts of interest.

\end{document}